# Strongly interacting bosons in a disordered optical lattice

M. White[1], M. Pasienski[1], D. McKay[1], S. Zhou[1], D. Ceperley[1], B. DeMarco[1]*

[1] Department of Physics, University of Illinois, 1110 W Green St, Urbana, IL 61801

*email: bdemarco@illinois.edu

Disorder, prevalent in nature, is intimately involved in such spectacular effects as the fractional quantum Hall effect and vortex pinning in type-II superconductors. Understanding the role of disorder is therefore of fundamental interest to materials research and condensed matter physics. Universal behavior, such as Anderson localization[1], in disordered non-interacting systems is well understood. But, the effects of disorder combined with strong interactions remains an outstanding challenge to theory. Here, we experimentally probe a paradigm for disordered, strongly-correlated bosonic systems—the disordered Bose-Hubbard (DBH) model—using a Bose-Einstein condensate (BEC) of ultra-cold atoms trapped in a completely characterized disordered optical lattice. We determine that disorder suppresses condensate fraction for superfluid (SF) or coexisting SF and Mott insulator (MI) phases by independently varying the disorder strength and the ratio of tunneling to interaction energy. In the future, these results can constrain theories of the DBH model and be extended to study disorder for strongly-correlated fermionic particles.

Despite being applied to many physical systems—from superfluids in porous media to superconducting thin films—questions remain regarding fundamental properties of the DBH model. For example, theoretical work using mean field theory[2-7], renormalization group[8], replica theory[9], and quantum Monte Carlo algorithms[10-15] disagree on the nature of the ground state phase diagram. While ultra-cold atom gases are ideal for studying disorder, having recently been applied to explore Anderson localization[16,17] and quasi-crystalline interacting 1D systems[18,19], the regime of strong interactions and disorder has not been previously accessed. To experimentally realize the DBH model, we add fine-grained disorder to an optical lattice potential using an optical speckle field with features as small as 570 nm. A unique aspect of this system compared with solid state materials is that the disorder strength is continuously tunable by controlling the intensity of the speckle field. We also use a measurement of the disordering potential to compute the single particle properties that parameterize the system, thereby completely characterizing the microscopic disorder.

We prepare BECs consisting of $(3.2\pm0.6)\times10^5$ atoms in a 3D cubic optical lattice potential with 406 nm between sites, as described in ref. 20. The lattice potential is characterized using a dimensionless parameter $s$, where the lattice potential depth is $sE_R$ ($E_R$ is the recoil energy at 812 nm). A parabolic confining potential generated by a magnetic trap and the Gaussian lattice beam profile leads to a range of site fillings, with approximately three atoms per site in the center of the lattice.

Fine-grained disorder is superimposed on the periodic optical lattice potential by passing 532 nm light through a holographic diffuser to generate an optical speckle field. The diffuser randomly scatters the light through a 0.5° range of angles, leading to a random distribution of intensities around the focal plane covering a Gaussian envelope with a 160 μm $1/e^2$ radius. Atoms experience a potential energy shift proportional to the intensity of the green laser light, leading to a combined potential that is a disordered lattice potential (Fig. 1b). The average potential energy shift $\Delta$ from the speckle field is used to quantify the disorder strength.

Several previous studies, beginning with ref. 21, have used speckle to experimentally investigate atoms confined in disordered potentials[16,17,22-24]. Compared with an earlier measurement in a 1D disordered optical lattice created using speckle[25], we have realized a new regime, in which the speckle size is comparable to the lattice spacing along every direction of a 3D lattice in the limit described by the DBH model. We achieve this by using relatively short-wavelength light, a high numerical aperture lens, and a geometry in which the speckle field propagates at 30° and 45° angles to the lattice directions (Fig.1a). The speckle length scale is characterized using the intensity autocorrelation (AC) function, which was determined by a measurement of the speckle intensity in a 10×100×80 µm volume using a scanning, high-resolution optical microscope (Fig. 1c). A fit to the measured AC distribution (Fig. 1d) to a Gaussian gives 790 and 650 nm $1/e^2$ radii, less than two lattice spacings, along the lattice directions. The shift in energy at nearest-neighbor lattice sites is therefore weakly correlated, which is typically assumed in theories of the DBH model.

At sufficiently low temperature and high *s*, atoms in a disordered lattice are described by the DBH Hamiltonian

$$H = \sum_i (\varepsilon_i - \tilde{\mu}_i) n_i - \sum_{\langle ij \rangle} t_{ij} b_i^\dagger b_j + \frac{1}{2} \sum_i U_i n_i (n_i - 1) \tag{1}$$

with site (kinetic and potential) energies $\varepsilon_i$, tunneling energies between nearest neighbors $t_{ij}$, on-site interaction energies $U_i$, and effective chemical potential $\tilde{\mu}_i = \mu - k r_i^2 / 2$ (µ is the chemical potential, *k* is the spring constant for the parabolic confining potential, and $r_i$ is the distance to the center of the lattice). The disordering potential gives rise to site-dependent distributions for the DBH model parameters ε, *t*, and *U*. The site energies are disordered by potential energy shifts, the tunneling energy by changes in the potential barrier between sites, and the interaction energy by modification of the curvature of the potential near the center of each lattice site. We define the tunneling and interaction energies as in ref. 26, although localized basis functions for each site must be used for the single-particle wavefunctions in the disordered potential.

In typical theoretical treatments of the DBH model, the disorder is limited to either the site or tunneling energies, which are generally assumed to be uniformly or Gaussian distributed. However, for our system, the statistical properties of the microscopic disorder are completely determined. We numerically compute the distributions for $\varepsilon$, $t$, and $U$ by calculating the local basis functions on a 3D disordered lattice potential with $7^3$ sites that reproduces the geometry shown in Fig. 1a; the parabolic confining potential is not included in this calculation.

Characteristic distributions are shown in Fig. 2a, b, and c. While the disorder results in distributions for $t$ and $U$ that are broadened to higher and lower values, the distribution for $\varepsilon$ is one-sided, since the blue-detuned speckle potential can only increase the potential energy of a lattice site. These distributions show that the disorder in $t$ and $\varepsilon$ is dominant compared with $U$—the relative scales of the disorder in each parameter are determined by the width of the distributions compared with their mean. As expected from the distribution of speckle intensities $I$, which is proportional to $e^{-I/\bar{I}}$ ($\bar{I}$ is the mean intensity), the width $\sigma_\varepsilon$ of the site energy distribution is proportional to $\Delta$. We have checked that the disorder does not shift the most probable values of $t$ and $U$, even though the mean values of these parameters change with increasing $\Delta$. The distributions shown in Fig. 2a confirm that the asymmetry of the speckle field with respect to the lattice directions does not play an important role in the distribution of tunneling energies. The distributions of tunneling energies along the $x$ and $z$ directions are not significantly different, and the average of the ratio of tunneling energy along the $x$ and $z$ direction does not deviate by more than 10% even for moderately strong disorder ($\Delta \sim U$).

Another aspect of the statistical properties of the DBH model is the correlations between the DBH model parameters. The correlation between $t$ and $\varepsilon$ is shown in scatter plots in Fig. 2d. While $t$ for a site averaged over all nearest neighbors is not strongly correlated to $\varepsilon$, $t_{ij}$ and the difference between site energies $|\varepsilon_i - \varepsilon_j|$ for nearest neighbors are directly correlated. This behavior is anticipated for fine-grained disorder, which induces local changes to the site energies coinciding with modifications to the potential barrier between neighboring sites.

In this work, we measure the effect of disorder on condensate fraction, which provides indirect information on the outstanding question of how the SF and MI phases in the uniform lattice transform when disorder is added. In their seminal paper on the DBH model[2], Fisher et al. argue that a Bose-glass (BG) phase intervenes between the MI and SF phases, implying that disorder transforms the MI phase into a BG phase. However, recent results using, for example, stochastic mean field[3] and replica theory[9] indicate that disorder can change the MI phase directly into a SF. We do not resolve this issue directly because condensate fraction is related, but not identical, to SF fraction. Only a portion of atoms are condensed because of the strong interactions in the DBH model; the condensate fraction is strictly zero for the MI phase and sufficiently localized BG phases.

We use time-of-flight imaging to measure the fraction of atoms in the condensate; a sample image and cross-sectional profile are shown as insets to Fig. 3c. We observe two-component profiles for all lattice depths $s$ and disorder strengths $\Delta$ used in this work, and define the number of atoms $N_0$ in the narrow central component as the condensate. We classify the number of atoms $N_{nc}$ in the broad component as the non-condensate, which may be atoms in MI or BG phases, or atoms in thermally excited states.

The measured condensate fraction $N_0/N=N_0/(N_0+N_{nc})$ is shown in Fig. 3a, b, and c for $s=6$, 12, and 14, three lattice depths that span different regimes for the pure Bose-Hubbard (BH) phase diagram. At zero temperature, the BH model with a parabolic confining potential predicts a SF phase with negligible quantum depletion at $s=6$, a strongly depleted SF phase at $s=12$, and coexisting MI and depleted SF phases at $s=14$ (refs 26,27). For the data in Fig. 3, the condensate fraction before transfer into the disordered lattice is greater than 90%. Data are shown for slow (15 ms) release from the lattice, which is used to measure the degree to which transfer into the disordered lattice is reversible, and fast (200 µs) release, which probes $N_0/N$ in the disordered lattice[28]. Ideally, in the absence of heating during transfer into the disordered lattice, the slow release would recover the condensate fraction prior to transfer.

We observe that disorder leads to partially reversible changes in condensate fraction. We interpret the irreversible decrease in $N_0/N$ measured for sufficiently large $\Delta$ as an increase in heating during transfer into the disordered lattice. This heating is never so strong that the condensate is destroyed; the lowest measured condensate fraction for slow release is 0.34 at $s=12$ and maximum $\Delta$. We were not able to determine the source of this heating, which may result from vibration of the speckle field relative to the lattice potential or long adiabatic timescales related to the disappearance of the band gap for strong disorder.

The significant reversible changes in $N_0/N$ evident in the fast release data in Fig. 3 may be caused by adiabatic transformations between phases induced by the disorder and by quantum depletion induced by interactions between atoms. Quantum depletion in the pure lattice at zero temperature is anticipated to be significant for $s=12$ and 14; site-decoupled mean field theory predicts $N_0/N$=0.97, 0.64, and 0.36 at $s=6$, 12, and 14 for our system (the ratio of atoms in the SF to MI phase is 0.83 at $s=14$)[29]. Because the disorder does not significantly affect the interaction energy, the systematic reduction of $N_0/N$ as $\Delta$ is increased for the fast release may be an indication that disorder is inducing a transformation between quantum phases.

To isolate the effect of the disordered lattice on condensate fraction from changes caused by heating, we define the reversible fractional change in $N_0/N$ as $\delta(N_0/N) = (N_0/N)_{fast}/(N_0/N)_{slow} - 1$, which is shown in Fig. 4 for each pair of slow and fast release data points in Fig. 3. If disorder results in no change in $N_0/N$ or in a change entirely due to heating, then $\delta(N_0/N) = 0$, which is indicated by the dashed line on Fig. 4. Several features are apparent in Fig. 4. An effect unrelated to disorder is the reduction in $N_0/N$ for $s=14$ and $\Delta=0$, which reflects the presence of atoms in the MI phase in the pure lattice. The reversible reduction in condensate fraction evident in Fig. 4 at all lattice depths is more pronounced at higher $s$ and saturates above $\Delta/U=15$ for $s=14$.

The inset to Fig. 4 shows an expanded view at low $\Delta$ for $s=14$. Data taken in this regime are sensitive to the effect of weak disorder on the unit filling MI phase. We observe that for relatively weak disorder at $s=14$, approximately a 10% reversible decrease in $N_0/N$ is measured up to $\Delta/U=1$ (with no change in $N_0/N$ after slow release). This result may seem to be at odds with predictions that disorder will transform the MI into SF in this regime[3,9]. However, before these data can constrain theory, finite temperature and the distribution of DBH model parameters shown in Fig. 2 must be included in calculations of condensate fraction. Furthermore, equation (1) requires corrections for multiple bands at the highest values of $\Delta$.

In conclusion, we have demonstrated that disorder in site and tunneling energy leads to depletion of condensate fraction in the DBH model (including parabolic confinement). The data in this manuscript may be used as a benchmark for predictions—$N_0/N$ can be straightforwardly computed in many theoretical approaches. In contrast to condensed matter systems, for which microscopic disorder cannot be fully characterized, we have completely computed the DBH model parameters for the disordered lattice, thereby eliminating the disorder as a free parameter in theory. The techniques we employ may also be used to study disorder in the (Fermi) Hubbard model, which is of interest to the high-temperature superconducting cuprates.

## METHODS

A 0.9 numerical aperture graded index lens (Lightpath Industries GPX-15-15) located approximately 13 mm from the atoms is used to focus the optical speckle field. The speckle intensity is calibrated by measuring the force from the envelope of the speckle field on a thermal gas. The Gaussian envelope of the speckle field gives rise to a repulsive potential that is used to deflect the center-of-mass of a 500 nK gas confined in the magnetic trap. By measuring this deflection as the alignment of the speckle field is changed, we are able to calibrate $\Delta$ with a 15% systematic uncertainty. We have confirmed that a 500 nK gas is sufficiently hot to eliminate effects from the fine-grained disorder by checking that the force from the speckle field envelope is linearly proportional to intensity.

After preparing a BEC in the magnetic trap, the 532 nm speckle field and 812 nm lattice light are turned on together over 100 ms using a sigmoidal ramp. Fast and slow release from the lattice is accomplished by turning down the lattice and speckle light linearly in 200 μs and 15 ms, respectively. The ratio $s/\Delta$ is kept fixed during the turn-on and turn-off of the disordered lattice.

In the numerical computation for the DBH model parameters, a basis set of single particle states $w_i(\vec{x} - \vec{x}_i)$ is constructed using imaginary time projection starting from states localized on the sites of the uniform lattice potential; $w_i(\vec{r}) = \langle \vec{r} | e^{-\tau[-\hbar^2 \vec{\nabla}^2/2m + V(\vec{r})]} | \vec{r}_i \rangle$, where $V$ is the disordered potential. The evolution is terminated when the high energy components are sufficiently suppressed but the basis functions are still well localized. The basis set is then orthogonalized using the Lowdin scheme in order to preserve the localization to the maximal extent[30]. The distributions of $\varepsilon_i = \int w_i^*(\vec{x} - \vec{x}_i)[-\hbar^2 \vec{\nabla}^2/2m + V(\vec{x})] w_i(\vec{x} - \vec{x}_i) d^3\vec{x}$, $t_{ij}$, and $U_i$ are calculated using this basis set, averaged over 150 random realizations of the speckle field (further details to be published Zhou and Ceperley). While the next-neighbor tunneling energies are strictly zero in the pure lattice, we find that disorder leads to finite next-neighbor tunneling energies (smaller by at least a factor of 1000 compared with the nearest-neighbor tunneling energies). A calculation of the density of states using the basis set constructed on the disordered lattice indicates that the gap between the ground and first excited band disappears for approximately $\Delta$=0.8, 1, and 1.75 $E_R$ (corresponding to $\Delta/U$=4.7, 3.3, and 5) for $s$=6, 12, and 14.

Images are taken after 20 ms time-of-flight and are fit to the combination of a Thomas-Fermi (TF) profile (for a harmonic trap) and Gaussian. The number of condensate atoms $N_0$ is determined from the TF component, and the number of non-condensate atoms $N_{nc}$ from the Gaussian. The width of the TF profile changes by less than 20% for the range of $\Delta$ in Fig. 3. Finite signal-to-noise ratio in our imaging system causes a systematic error for measurements of low $N_0/N$ at $s$=12 and 14. We use an empirical imaging noise model to determine this systematic error, which is less than 0.03 in $N_0/N$ for the data in this paper.

# References

[1] Anderson, P.W., Absence of diffusion in certain random lattices. *Phys. Rev.* **109**, 1492-1505 (1958).

[2] Fisher, M.P., Weichman, P.B., Grinstein, G., and Fisher, D. S., Boson localization and the superfluid-insulator transition. *Phys. Rev. B* **40**, 546–570 (1989).

[3] Bissbort, U. and Hofstetter, W., Stochastic mean-field theory for the disordered Bose-Hubbard model. *arxiv:0804.0007* (2008).

[4] Buonsante, P., Penna, V., Vezzani, A., and Blakie, P. B., Mean-field phase diagram of cold lattice bosons in disordered potentials. *Phys. Rev. A* **76**, 011602(R) (2007).

[5] Damski, B. et al., Atomic Bose and Anderson glasses in optical lattices. *Phys. Rev. Lett.* **91**, 080403 (2003).

[6] Sheshadri, K., Krishnamurthy, H. R., Pandit, R., and Ramakrishnan, T. V., Percolation-enhanced localization in the disordered bosonic Hubbard-model. *Phys. Rev. Lett.* **75**, 4075-4078 (1995).

[7] Krutitsky, K. V., Pelster, A., and Graham, R., Mean-field phase diagram of disordered bosons in a lattice at nonzero temperature. *New J. Phys.* **8**, 187 (2006).

[8] Singh, K. G. and Rokhsar, D. S., Real-space renormalization study of disordered interacting bosons. *Phys. Rev. B* **46**, 3002-3008 (1992).

[9] Wu, J. and Phillips, P., Disorder-induced missing moment of inertia in solid 4He. *arxiv:cond-mat/0612505* (2007).

[10] Balabanyan, K. G., Prokof'ev, N., and Svistunov, B., Superfluid-insulator transition in a commensurate one-dimensional bosonic system with off-diagonal disorder. *Phys. Rev. Lett.* **95**, 055701 (2005).

[11] Prokof'ev, N. and Svistunov, B., Superfluid-insulator transition in commensurate disordered bosonic systems: Large-scale worm algorithm simulations. *Phys. Rev. Lett.* **92**, 015703 (2004).

[12] Krauth, W., Trivedi, N., and Ceperley, D., Superfluid-insulator transition in disordered boson systems. *Phys. Rev. Lett.* **67**, 2307-2310 (1991).

[13] Kisker, J. and Rieger, H., Bose-glass and Mott-insulator phase in the disordered boson Hubbard model. *Phys. Rev. B* **55**, 11981-11984 (1997).

[14] Sengupta, P., Raghavan, A., and Haas, S., Disorder-enhanced phase coherence in trapped bosons on optical lattices. *New J. Phys.* **9**, 103 (2007).

[15] Scalettar, R. T., Batrouni, G. G., and Zimanyi, G. T., Localization in interacting, disordered, Bose systems. *Phys. Rev. Lett.* **66**, 3144-3147 (1991).

[16] Roati, G. et al., Anderson localization of a non-interacting Bose-Einstein condensate. *Nature* **453**, 895-898 (2008).

[17] Billy, J. et al., Direct observation of Anderson localization of matter waves in a controlled disorder. *Nature* **453**, 891-894 (2008).

[18] Fallani, L. et al., Ultracold atoms in a disordered crystal of light: Towards a Bose glass. *Phys. Rev. Lett.* **98**, 130404 (2007).

[19] Lye, J. E. et al., Effect of interactions on the localization of a Bose-Einstein condensate in a quasiperiodic lattice. *Phys. Rev. A* **75**, 061603(R) (2007).

[20] McKay, D., White, M., Pasienski, M., and DeMarco, B., Phase-slip-induced dissipation in an atomic Bose-Hubbard system. *Nature* **453**, 76-79 (2008).

[21] Lye, J. E. et al., Bose-Einstein condensate in a random potential. *Phys. Rev. Lett.* **95**, 070401 (2005).

[22] Clement, D. et al., Experimental study of the transport of coherent interacting matter-waves in a 1D random potential induced by laser speckle. *New J. Phys.* **8**, 165 (2006).

[23] Chen, Y. P. et al., Phase coherence and superfluid-insulator transition in a disordered Bose-Einstein condensate. *Phys. Rev. A* **77**, 033632 (2008).

[24] Clement, D., Bouyer, P., Aspect, A., and Sanchez-Palencia, L., Density modulations in an elongated Bose-Einstein condensate released from a disordered potential. *Phys. Rev. A* **77**, - (2008).

[25] Schulte, T. et al., Routes towards Anderson-like localization of Bose-Einstein condensates in disordered optical lattices. *Phys. Rev. Lett.* **95**, 170411 (2005).

[26] Jaksch, D. et al., Cold bosonic atoms in optical lattices. *Phys. Rev. Lett.* **81**, 3108-3111 (1998).

[27] DeMarco, B., Lannert, C., Vishveshwara, S., and Wei, T.-C., Structure and stability of Mott-insulator shells of bosons trapped in an optical lattice. *Phys. Rev. A* **71**, 063601 (2005).

[28] Greiner, M. et al., Exploring phase coherence in a 2D lattice of Bose-Einstein condensates. *Phys. Rev. Lett.* **87**, 160405 (2001).

[29] Sheshadri, K., Krishnamurthy, H. R., Pandit, R., and Ramakrishnan, T. V., Superfluid and insulating phases in an interacting-boson model: Mean-field theory and the RPA. *Europhys. Lett.* **22**, 257-263 (1993).

[30] Aiken, J.G., Erdos, J.A., and Goldstein, J.A., On Lowdin orthogonalization. *Int. J. Quantum Chem.* **18**, 1101-1108 (1980).

**Acknowledgements**

The authors thank W. Hofstetter, N. Trivedi, and P. Phillips for stimulating discussions. This work was funded by the Army Research Office (award W911NF-08-1-0021) and the National Science Foundation (award 0448354). Any opinions, findings, and conclusions or recommendations expressed in this material are those of the authors and do not necessarily reflect the views of the National Science Foundation. B. DeMarco acknowledges support from the Sloan Foundation, and D. McKay recognizes support from NSERC.

**Figure 1. Fine-grained disorder superimposed on a 3D periodic optical lattice potential. a,** A polycarbonate holographic diffuser (gray) from Luminit LLC is used to create a 532 nm optical speckle field. The light (green) is focused by a high numerical aperture lens into the rectangular glass cell, where a BEC (blue circle) is confined in a three-dimensional optical lattice formed from three pairs of counter-propagating 812 nm laser beams (red). **b,** The random optical potential (green) created by the optical speckle field adds independently to the lattice potential (red) to create a disordered potential. The average Stark shift $\Delta$ from the disordering potential is used to characterize the disorder strength. **c,** A sample of the volume of the speckle intensity distribution (shown in false color) measured using a microscope. **d,** The measured speckle intensity is used to reconstruct the AC function, which is shown in false color. The $1/e^2$ radii of the AC distribution are 570 nm and 3 µm along the transverse and speckle propagation directions; the lattice axes project onto the $x$ and $z$ directions.

**Figure 2. Distributions of the DBH model parameters for $s=14$ and $\Delta=1$ $E_R$. a, b, c,** Sample distributions of DBH parameters. **a,** The inset shows the dependence of the standard deviation $\sigma_\varepsilon$ of the site energy distribution on $\Delta$; the line is a fit with slope $0.838 \pm 0.006$. **b,** The tunneling energy distributions are computed separately along the $x$ (blue) and $z$ (green) directions. The inset shows the average ratio of the tunneling energy along the $x$ and $z$ directions. **d,** Correlations between tunneling and site energies. A scatter plot of $t_{ij}$, the tunneling energy between neighboring sites $i$ and $j$, and the difference in site energies $|\varepsilon_i - \varepsilon_j|$ exhibits a direct correlation. The inset is a scatter plot of the tunneling energy averaged along the lattice directions ($t_i$) vs. the site energy for each site.

**Figure 3. Measured condensate fraction $N_0/N$ in the disordered lattice potential.** Data are shown for **a,** $s=6$, **b,** $s=12$, and **c,** $s=14$. The disorder strength is normalized to $U$ for each $s$; the range of disorder strength corresponds to $\Delta=0$–$7$ $E_R$ for **a** and **b**, and $\Delta=0$–$6$ $E_R$ for **c**. The hollow points are the condensate fraction after slow release from the lattice, and the solid points after fast release. Each data point is the average of more than 7 measurements at fixed $\Delta$ and $s$, with the error bars showing the statistical uncertainty. The inset to **c** is an image with a 400 µm field-of-view taken at $s=14$ and $\Delta=2$ $E_R$, corresponding $\Delta/U=5.7$. The cross-sectional profile of this image shows the fit to a two-component distribution (solid line) used to determine $N_0/N$.

**Figure 4. The reversible change in condensate fraction induced by disorder.** Data are shown for $s=6$ (green), $s=12$ (red), and $s=14$ (blue). Each data point is derived from a pair of slow and fast release points from Fig. 3; the errors bars are computed using the statistical uncertainties from Fig. 3. The inset is an expanded view of the low $\Delta/U$ data for $s=14$.

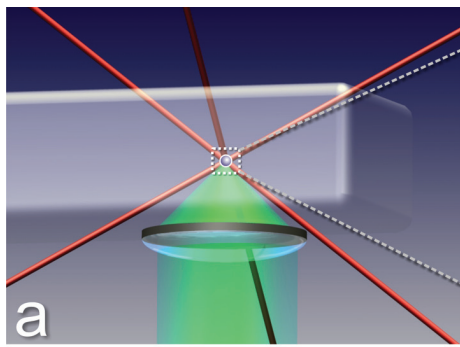
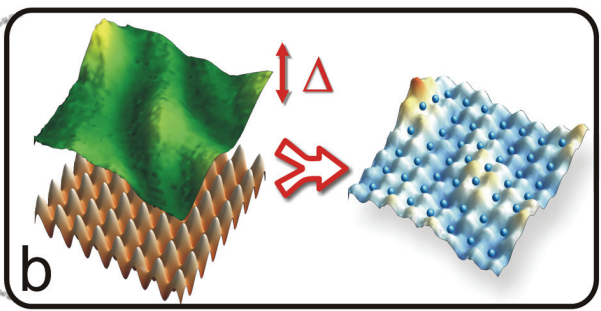
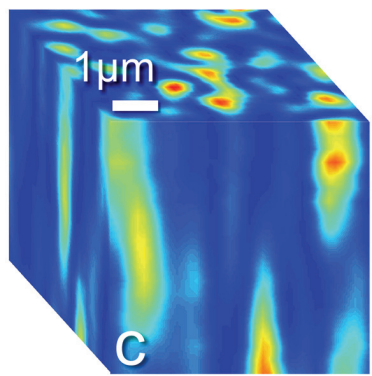
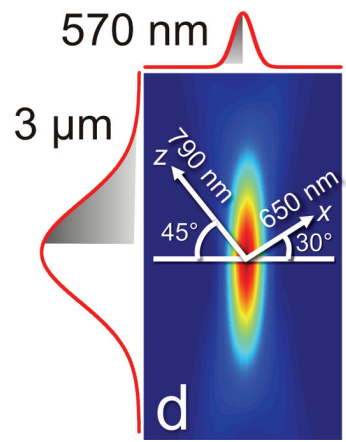

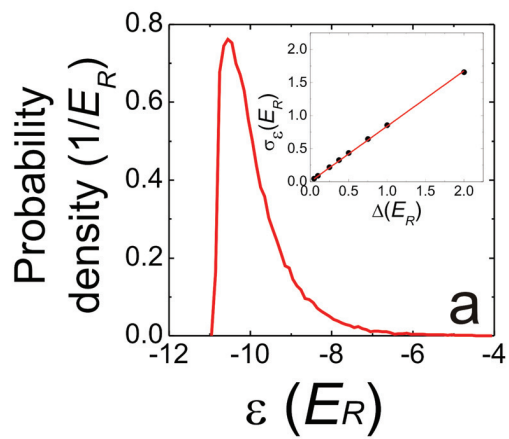
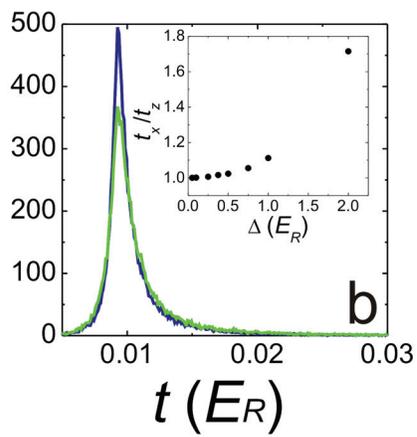
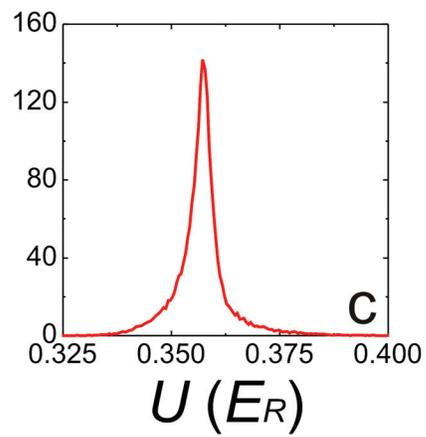
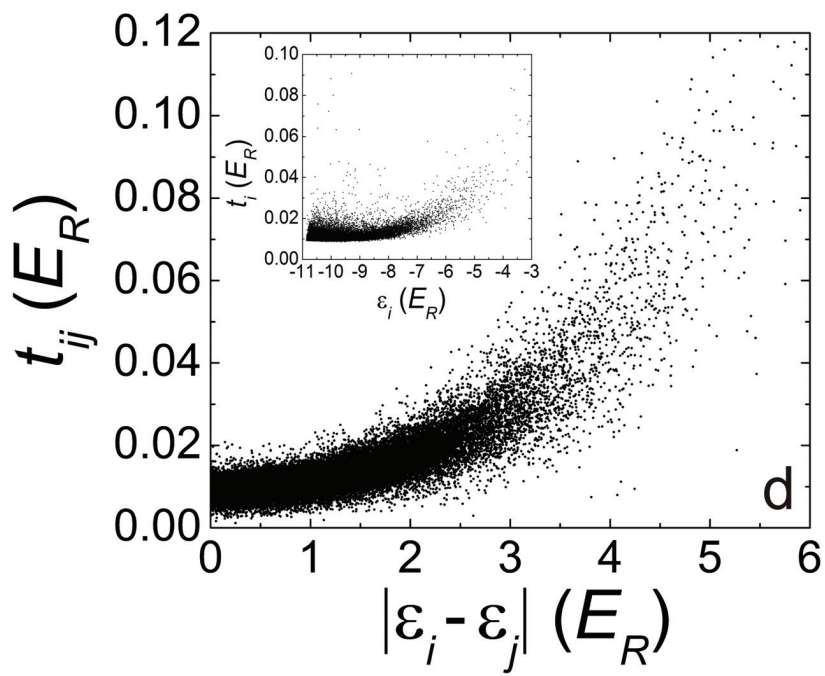

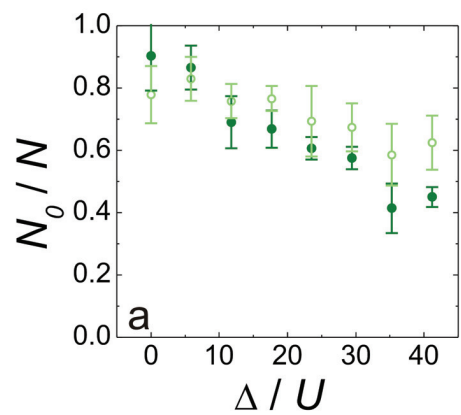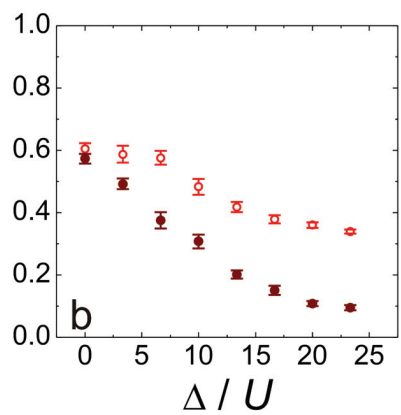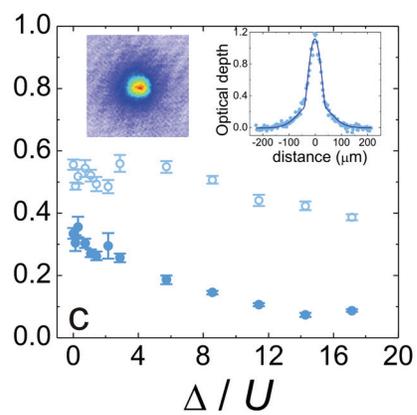

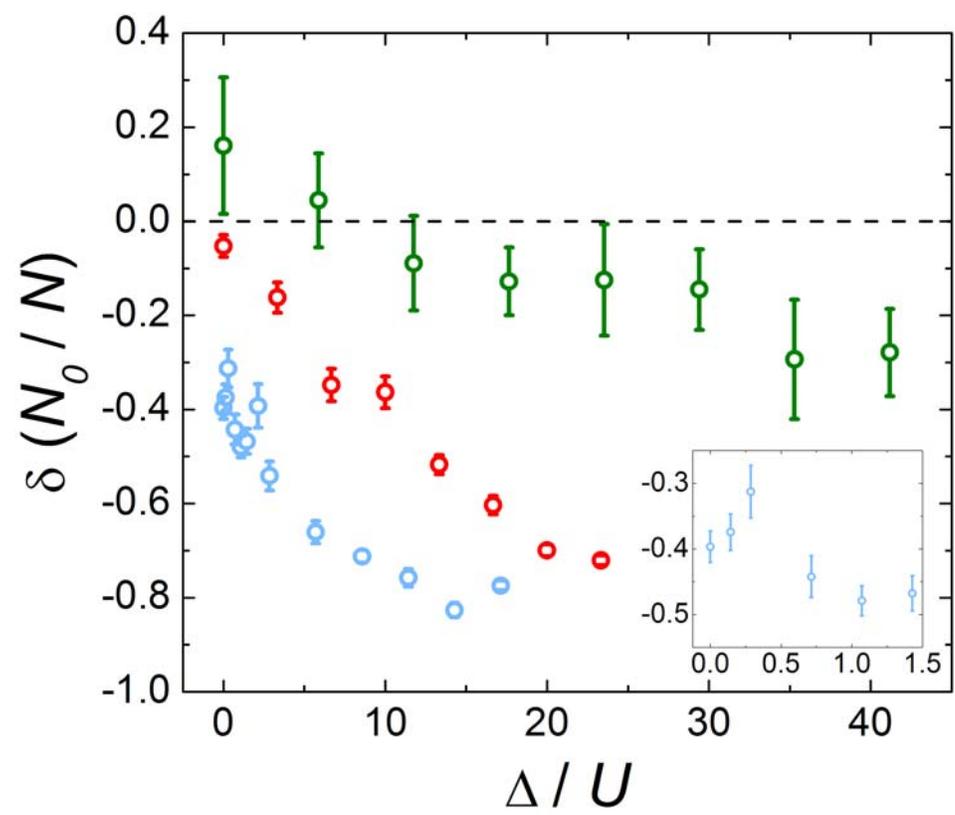